\documentstyle[aps,twocolumn]{revtex}
\begin{document}
\voffset 0.5in
\draft
\wideabs{
\title{Nonuniqueness and derivative discontinuities in density-functional
theories for current-carrying and superconducting systems}
\author{K. Capelle}
\address{Instituto de Qu\'{\i}mica de S\~ao Carlos,
Universidade de S\~ao Paulo, Caixa Postal 780, S\~ao Carlos, 13560-970 SP,
Brazil}
\author{G. Vignale}
\address{
Department of Physics and Astronomy, University of Missouri-Columbia,
Columbia, Missouri 65211, USA}
\date{\today}
\maketitle
\begin{abstract}
Current-carrying and superconducting systems can be treated within
density-functional theory if suitable additional density variables
(the current density and the superconducting order parameter, respectively)
are included in the density-functional formalism. Here we show that
the corresponding conjugate potentials (vector and pair potentials,
respectively) are {\it not} uniquely determined by the densities.
The Hohenberg-Kohn theorem of these generalized density-functional theories
is thus weaker than the original one. We give explicit examples and explore
some consequences.
\end{abstract}
}
\newcommand{\be}{\begin{equation}}
\newcommand{\ee}{\end{equation}}
\newcommand{\bea}{\begin{eqnarray}}
\newcommand{\eea}{\end{eqnarray}}
\newcommand{\bi}{\bibitem}

\newcommand{\ep}{\epsilon}
\newcommand{\s}{\sigma}
\newcommand{\p}{{\bf \pi}}
\newcommand{\D}{\Delta}
\newcommand{\r}{({\bf r})}
\newcommand{\rp}{({\bf r'})}
\newcommand{\rrp}{({\bf r},{\bf r'})}
\newcommand{\xR}{({\bf x},{\bf R})}

\newcommand{\ua}{\uparrow}
\newcommand{\da}{\downarrow}
\newcommand{\la}{\langle}
\newcommand{\ra}{\rangle}
\newcommand{\dg}{\dagger}

\newcommand{\lmt}{\left(\begin{array}{cc}}
\newcommand{\lmf}{\left(\begin{array}{cccc}}
\newcommand{\rmat}{\end{array}\right)}
\newcommand{\lvec}{\left(\begin{array}{c}}
\newcommand{\rvec}{\end{array}\right)}

Today, density-functional theory (DFT) \cite{kohnrmp} is an indispensable
tool for the investigation of the electronic structure of matter
in atomic, molecular, or extended systems.
The theory rests on the celebrated Hohenberg-Kohn (HK) theorem \cite{hk},
which guarantees that the (v-representable) ground-state density
$n\r$ uniquely determines the ground-state many-body wave function
$\psi_0({\bf r}_1,\ldots,{\bf r}_N)$. This theorem on its own is a very
powerful result, but in the original formulation of DFT \cite{hk,ks} one can 
prove even more: the external potential $v\r$ (e.g., the nuclear charge 
distribution in a molecule or a solid),
too, is a functional of the density, and is unique up to an additive constant.
Since this external potential in turn determines all eigenstates of the
many-body Hamiltonian, this implies that {\it all} observables (and not
only ground-state ones) are functionals of the ground-state density.

Following original ideas of von Barth and Hedin \cite{vbh}, it has
recently been shown by Eschrig and Pickett \cite{ep} and by the present
authors \cite{uniqueprl} that in spin-DFT (SDFT) the situation is not
that simple: while the wave function is still uniquely determined by
the spin densities $n_\ua\r$ and $n_\da\r$, the external potentials
$v_\ua\r$ and $v_\da\r$ [or $v\r$ and ${\bf B}\r$] are not. This implies
that SDFT functionals are not always differentiable, and has far-reaching 
consequences for the construction of better exchange-correlation (xc)
functionals, and for applications to systems such as half-metallic
ferromagnets \cite{ep,uniqueprl}.

SDFT is not the only instance at which the original HK theorem has
been generalized. In the present work we extend the analysis of
Ref.~\cite{uniqueprl} to two other generalizations of DFT, namely
current-DFT (CDFT) \cite{vr1,vr2}.
and DFT for superconductors \cite{ogk1,ogk2,ldaprl,balazsprl}.
The discovery of nonuniqueness in these generalized DFTs deepens our
understanding of the respective xc functionals and flags a warning
signal to all-to-immediate generalizations of the original
HK theorem to more complex situations.

The basic physics of nonuniqueness is simple. When a sufficiently small
change in one of the external fields does not change the corresponding 
density distribution, the corresponding susceptibility vanishes. 
The search for, and the interpretation of, nonuniqueness in DFT is 
thus guided by investigations of the circumstances under which some 
response function becomes zero. 

We first consider current-carrying systems.
The appropriate formulation of (nonrelativistic) DFT is CDFT
\cite{vr1,vr2}, which is based on the many-body Hamiltonian 
(in atomic units, i.e., $\hbar=e=m=1$)
\bea
\hat{H}=\hat{T}+\hat{U}+
\int d^3r\, \hat{n}\r [v\r -\mu]
+\frac{1}{c}\int d^3r \, \hat{{\bf j}}_p\r {\bf A}\r
\nonumber \\
+ \frac{1}{2c^2}\int d^3r\, \hat{n}\r A^2\r
+\int d^3r \, \hat{{\bf m}}\r {\bf B}\r,
\label{cham}
\eea
where ${\bf B}\r=\nabla\times{\bf A}\r$ is the magnetic field,
$v\r$ the electrostatic one, and $\hat{T}$ and $\hat{U}$ denote the operators
for kinetic energy and particle-particle interaction, respectively.

The basic variables of CDFT, in terms of which the entire ground-state
physics of the current-carrying many-body system is described, are 
$n\r$, ${\bf m}\r$, and ${\bf j}_p\r$, the ground-state expectation values 
of the particle density operator 
$\hat{n}\r =\sum_\sigma\Psi_\sigma^\dagger\r \Psi_\sigma\r $,
spin magnetization operator $\hat{\bf m}\r=
(1/2c)\sum_{\alpha,\beta}\Psi_\alpha\r\hat{\bbox{\sigma}}\Psi_\beta\r$,
and (paramagnetic) current density operator
\be
\hat{{\bf j}}_p\r 
=\frac{1}{2i}\sum_\sigma \left[
\Psi_\sigma^\dagger\r(\nabla\Psi_\sigma\r)-
(\nabla\Psi_\sigma^\dagger\r)\Psi_\sigma\r
\right],
\ee
where the $\Psi_\sigma\r$ are field operators and $\hat{\bbox{\sigma}}$
is the vector of Pauli matrices.

According to the CDFT version of the HK theorem these densities
uniquely determine the ground-state many-body wave function. However, in 
striking contrast to conventional `density-only' DFT they do {\it not} 
uniquely determine the potentials ${\bf A}\r$, ${\bf B}\r$, and $v\r$: it
is possible to find different vector and scalar potentials which yield
the same ground state, and consequently the same densities
$n\r$, ${\bf m}\r$, and ${\bf j}_p\r$.

Before delving into a general characterization of such potentials,
we present a simple example that clearly displays the problem.  
We consider an atom subjected to a uniform magnetic field 
${\bf B}=B\hat{\bf z}$, where $\hat{\bf z}$ is the unit vector along the 
$z$-axis. Ignoring spin-orbit
interactions, the Hamiltonian can be written as 
\be
\label{Hatom}
\hat H = \sum_i \left ( {\hat p_i^2 \over 2}-{Z \over r_i} + 
\frac{B^2 r_{\perp,i}^2}{8c^2} \right ) 
+ \sum_{i \neq j} \frac{1}{2r_{ij}}
+ \frac{\hat L_z + 2 \hat S_z}{2c} B,
\ee
where $r_{\perp,i}^2 \equiv x_i^2+ y_i^2$, and $\hat L_z$ and $\hat S_z$ are 
the $z$-components of the orbital and spin angular momentum operators 
${\bf L}$ and ${\bf S}$. $Z$ is the atomic number, specifying the
external potential. 
Both $\hat L_z$ and $\hat S_z$ are constants of 
motion, hence the (nondegenerate) ground state of $\hat H$
is also an eigenstate of $\hat L_z$ and $\hat S_z$, with eigenvalues 
$m_L$ and $m_S$ respectively.  The ground-state energy is $E_0$.

Consider now the same system of $Z$ electrons being subjected to a different
(but still uniform) magnetic field ${\bf B'}=B'\hat{\bf z}$ and the
external potential
\begin{equation}
\label{Vprimeatom}
v'(r) = -{Z \over r} - {1\over 8 c^2} (B'^2 - B^2) r_{\perp}^2.
\end{equation}
The Hamiltonian of this system is
\be
\label{Hprime}
\hat H' 
= \hat H + {1 \over 2c} (\hat L_z + 2 \hat S_z) (B' - B).
\ee

Thus, we immediately see that the ground state of $\hat H$ (or, for that
matter, any simultaneous eigenstate of $\hat H$, $\hat L_z$, and $\hat S_z$)
is also an eigenstate of $\hat H'$ with eigenvalue 
$E' = E_0 + (1/2c)(m_L+2m_S) (B'-B)$.  
Furthermore, if the difference $B'-B$ is not
too large, this eigenstate will be the ground state of $\hat H'$;  the
qualitative  condition for this to happen is that $E'-E_0 << E_G$
where $E_G$ is the energy gap between the first excited state and the
ground state of $\hat H$.
Thus, we have succeeded in constructing two different sets of potentials,
${\bf A} = ({\bf B} \times {\bf r})/2$ and $v$, and
${\bf A'} = ({\bf B'} \times {\bf r})/2$ and $v'$, 
that yield the same ground state. 

Let us now consider the question from a more general point of view.  Let
${\bf A'} = {\bf A} + \Delta {\bf A}$ and $v' = v + \Delta v$ be vector and
scalar  potentials that are supposed to yield the same ground state
$\psi_0$ as ${\bf A}$ and $v$. A necessary condition for this 
is that $\psi_0$ satisfy the eigenvalue equation
\be
\label{eigenvalueproblem}
\int  d^3r \left[\hat{n}\Delta v +{1 \over 2c^2} \hat{n} \Delta {\bf A}^2 
+{1\over c} \hat{\bf j}_p \Delta {\bf A} \right] \psi_0
= \Delta E \psi_0,
\ee
where we neglected, for simplicity, the spin-degrees of freedom, because 
the nonuniqueness associated with them is already discussed in Refs.~\cite{ep} 
and \cite{uniqueprl}.

The general problem at hand is thus to find a linear
combination of the density operators $\hat n\r$ and $\hat{\bf j}_p\r$ 
that has $\psi_0$ as eigenfunction. This problem is not easily
solved in general. It is easy, however, to obtain a particular solution of
Eq.~(\ref{eigenvalueproblem}) if one can find a linear combination of the
density operators that is a constant of motion. The
ground state of $\hat H$ is automatically an eigenstate of such a constant
of motion, and Eq.~(\ref{eigenvalueproblem}) is satisfied. By making the
coefficients of the linear combination sufficiently small we can always ensure
that $\psi_0$ remains the ground state of the Hamiltonian with the new
potentials (assuming of course that the spectrum of $\hat H$ has a
gap between its ground state and first excited state).
This is the same prescription employed in Ref. \cite{uniqueprl} to construct
examples for nonuniqueness in SDFT. In the terminology of that reference
nonuniqueness arising from such constants of motion is referred to
as {\it systematic nonuniqueness}.

As a trivial example of this procedure consider the constant of motion
$ \hat N = \int d^3r\, \hat n\r$.
The existence of this constant of motion tells us that $\Delta v\r =
const$, $\Delta {\bf A}\r = 0 $ is a solution of
Eq. (\ref{eigenvalueproblem}). This is the well known nonuniqueness of the
scalar potential with respect to the addition of a constant.
Consider now the less trivial example
\begin {equation}
\label{constant2}
\hat L_z =  \int d^3r \,({\bf \hat z} \times {\bf r}) \cdot \hat{\bf j}_p\r,
\end{equation}
which is a constant of motion in any system that is  invariant
under rotations about the z-axis.
Comparing this with Eq.~(\ref{eigenvalueproblem}) we immediately see that
$\Delta {\bf A}\r = \Delta B ({\bf \hat z} \times {\bf r})/2$
and
$\Delta v\r =  -[\Delta {\bf A}\r]^2/(2c^2)$
with $\Delta B=const$ is indeed a solution of the posed problem.  This
is, of course, nothing but a more formal derivation of the elementary example 
discussed above.

Another way in which nonuniqueness can arise is by adding an operator to the 
Hamiltonian which, although not a constant of motion, happens to have 
eigenvalue zero on the ground state. This was
called {\it accidental nonuniqueness} in Ref. \cite{uniqueprl}.
To give an example in CDFT, let $n_\s\r$ and ${\bf j}_{p\s}\r$ denote the 
exact spin-resolved ground-state density and paramagnetic current of a 
two-electron system, such as the $He$ atom, in the presence of external 
vector and scalar potentials. For sufficiently small external fields
these densities must arise from the single particle orbitals 
$\varphi_\ua\r$ and $\varphi_\da\r$ that are the lowest energy solutions of 
the spin-dependent Kohn-Sham (KS) equations
\begin{equation}
\label {KS1}
\left\{ {1 \over 2} \left ( - i {\bf \nabla} + {1 \over c} {\bf A}_{s\s}
\r \right )^2 + v_{s\s} \r \right \} \varphi_\s\r = \epsilon_\s \varphi_\s\r,
 \end {equation}
where ${\bf A}_{s\s} \r$ and $v_{s\s} \r$ are the KS potentials,
defined, as usual, in terms of the external, Hartree, and exchange-correlation 
potentials.
The relation between the densities and the single-particle orbitals is
$n_\s\r = |\varphi_\s\r |^2$
and
\be
{\bf j}_{p\s} \r = n_\s\r {\bf \nabla} \phi_\s\r,
\label{jorbital}
\ee
where $\phi_\s\r$ is the phase of the complex orbital $\varphi_\s\r$
\cite{vorticityfootnote}.

Eq.~(\ref{KS1}) can be rewritten in the form
\be
\left \{ {1 \over 2} \left ( - i {\bf \nabla} + {1 \over c} [{\bf A}_{s\s}
 + c {\bf \nabla} \phi_\s  ]\right )^2
+ v_{s\s}  \right \}
|\varphi_\s|
= \epsilon_\s |\varphi_\s |,
\label {KS2}
\ee
which has the solution
${\bf A}_{s\s}\r = -c{\bf \nabla} \phi_\s\r = -c{\bf j}_{p\s}\r/n_\s\r$ and
\be
v_{s\s} \r = 
{1\over 2} {\nabla^2 |\varphi_\s \r| \over |\varphi_\s \r|} +\epsilon_\s
= {1\over 2} {\nabla^2 n_\s^{1/2} \r \over n_\s^{1/2} \r} +\epsilon_\s.
\ee

To determine whether these are the only potentials that reproduce the given 
densities $n_\s\r$ and ${\bf j}_{p\s}\r$ we assume the
existence of a second such set of potentials,
${\bf A}'_{s\s} \r = {\bf A}_{s\s} \r + \Delta {\bf A}_{s\s}\r$
and
$v_\s\r =v_\s\r + \Delta v_{s\s}\r$. 
By substituting these back in Eq.~(\ref{KS2})
and separating the real and the imaginary parts we obtain
\begin{equation}
\label{KS3}
v'_{s\s} \r + {1\over 2 c^2}\Delta{\bf A}^2_{s\s} \r   =  {1\over 2 }
{\nabla^2 n_\s^{1/2} \r \over n_\s^{1/2} \r} + \epsilon_\s
\end{equation}
and $ {\bf \nabla} \cdot [n_\s\r \Delta{\bf A}_{s\s} \r] = 0$.
This last equation follows more directly from the application of the continuity
equation to the real solution of (\ref{KS2}). Its general solution is
$\Delta{\bf A}_{s\s} \r = {\bf \nabla} \times {\bf Q}_\s\r / n_\s\r$
where ${\bf Q}_\s\r$ is an arbitrary vector field. Hence,
\begin{equation}
\label{solvks}
v'_{s\s} \r   =  v_{s\s}\r
 - {1\over 2 c^2} \left ( {{\bf \nabla} \times {\bf Q}_\s\r
\over n_\s \r} \right )^2
\end{equation}
and
\begin{equation}
\label {solA}
{\bf A}'_{s\s} \r = {\bf A}_{s\s}\r
+ {  {\bf \nabla} \times {\bf Q}_\s \r \over n_\s \r}.
\end{equation}
By construction, ${\bf A}_{s\s}\r$ and $v_{s\s}\r$ are the potentials for 
which $n_\s\r$ and ${\bf j}_{p\s}\r$ are {\it ground-state} densities. 
If ${\bf Q}_\s$ is sufficiently small and if the KS system at 
${\bf Q}_\s=0$ has an energy gap separating the first excited state from the
ground state, $\varphi_\s\r$ will remain the ground state in the
potentials ${\bf A}_{s\s}'\r$ and $v_{s\s}'\r$.
Thus, Eqs. (\ref{solvks}) and (\ref{solA}) provide a vivid and nontrivial
example of nonuniqueness of the KS potentials of CDFT. 

Next, we turn to the superconducting case. Here the underlying many-body 
Hamiltonian is \cite{ogk1}
\bea
\hat{H} =
\hat{T}+\hat{U} + \int d^3r\,\hat{n}\r  [v\r -\mu]
+\int d^3r\, \hat{\bf m}\r {\bf B}\r
\nonumber \\
-\int d^3r \int d^3r' [\hat{\chi}\rrp D^*\rrp + H.c.], 
\label{SCham}
\eea
where the expectation value of the pair operator 
$\hat{\chi}\rrp = \Psi_\ua\r \Psi_\da\rp$ is the superconducting order 
parameter and $D\rrp$ the corresponding pair potential. 
The phonon-induced
interaction term of Ref.~\cite{ogk1} can be added to $\hat{H}$ without 
changing our conclusions. 

As above, we now assume that the densities $n\r$, ${\bf m}\r$ and
$\chi\rrp$ can also be reproduced in different fields $v'=v+\Delta v$,
${\bf B}'={\bf B}+\Delta {\bf B}$ and $D' = D + \Delta D$. 
The equation obeyed by $\Delta v$, $\Delta {\bf B}$ and $\Delta D$ is
\bea
\int d^3r \left[\hat{n} \Delta v + {\bf \hat{m}} \Delta{\bf B}
-  \int d^3r' (\hat{\chi}\Delta D + H.c.)
\right] \psi_0 
\nonumber \\ =\Delta E \psi_0.
\label{SCdiff}
\eea

At this stage we already see a first nontrivial difference to the case of
CDFT and SDFT:
due to the presence of the pair operator $\hat{\chi}$ in
$\hat{H}$ the particle number operator $\hat{N}$ is not a constant of motion,
and we are {\it not} free to add an arbitrary
constant to the external potential $v\r$. In other words
$\Delta D=0$, $\Delta {\bf B}=0$ and $\Delta v =const$ 
is not a solution of Eq.~(\ref{SCdiff}) for a given $\psi_0$.
DFT for superconductors thus does not suffer from the most basic 
nonuniqueness of all, that with respect to the additive constant in the 
electrostatic potential.

However, DFT for superconductors is not free of nonuniqueness. 
For a singlet superconductor the spin susceptibility 
vanishes at zero temperature \cite{tinkham}. In the light of our physical 
characterization of nonuniqueness at the beginning of this paper we would thus 
expect some associated nonuniqueness. Indeed, this is bourne out by more 
detailed analysis. 
If ${\bf B}=B\hat{\bf z}$ is spatially uniform and sufficiently weak not to 
break Cooper pairs paramagnetically, then ${\bf B}'={\bf B}+\Delta {\bf B}$,
where $\Delta {\bf B}$ is also weak, uniform and parallel to $\hat{\bf z}$, 
has the same ground state, because under these circumstances
$\hat{M}_z$ = $\int d^3r\,\hat{m}_z$ is a conserved quantity, i.e., the
superconductor remains in a singlet state, with all electrons paired up.
Consequently, the set of potentials $\{v,{\bf B},D\}$ is not uniquely
determined by the conjugate densities $\{n,{\bf m},\chi\}$.
Since it is associated with the constant of motion $\hat{M}_z$, this is
{\it systematic nonuniqueness} in the above sense.

With these examples we end our list of explicit occurences of
nonuniqueness in generalized DFTs, and now turn  
to a discussion of broader aspects of our findings. 

In early papers on both CDFT\cite{vr1,vr2} and DFT for superconductors
\cite{ogk2} one finds the statement that the chosen densities uniquely
determine the corresponding potentials. As we have shown here, these 
statements are not accurate, and all that is determined uniquely is the 
ground-state wave function. Concerning consequences of this finding we refer 
the reader to the discussion we have given earlier of consequences of 
nonuniqueness in SDFT \cite{uniqueprl}. That discussion carries over 
almost literally to the case of current-carrying and superconducting systems. 
However, we wish to stress particularly that for most applications
of any DFT, including CDFT and DFT for superconductors, uniqueness of the 
ground-state wave function is sufficient, since no explicit use of the 
density-potential relation is made.  A notable exception within CDFT is the 
recent work by Handy and Lee \cite{handylee},
in which it is attempted to systematically construct exact CDFT potentials
from given densities. This construction must be reexamined in view of our 
finding that the CDFT potentials are not uniquely determined by the 
densities. Further exceptions are listed in Ref.~\cite{uniqueprl}.

Another important consequence of non\-unique\-ness in
DFT arises from the connection between the external and 
KS potentials with the functional derivatives of the kinetic and
internal energy functionals. For CDFT these connections take the form
(neglecting, for simplicity, again the spin degrees of freedom)
\be
-\frac{\delta T_s[n,{\bf j}_p]}{\delta n\r} 
= v_s\r - \mu + \frac{1}{2c} {\bf A}_s\r^2
\label{ts1}
\ee
and 
\be
-\frac{\delta T_s[n,{\bf j}_p]}{\delta {\bf j}_p\r}
= \frac{1}{c} {\bf A}_s\r,
\label{ts2}
\ee
i.e., derivatives of the noninteracting kinetic energy $T_s$ determine
the KS potentials $v_s$ and ${\bf A}_s$.
Similarly, the derivatives of the internal energy functional
$F$, defined as the ground-state expectation value of $\hat{T}$ and
$\hat{U}$, determine the external potentials $v\r$ and ${\bf A}\r$.
Analogous equations hold also in DFT for superconductors.

From Eqs.~(\ref{ts1}) and (\ref{ts2}) we
see that nonuniqueness of the Kohn-Sham potentials implies that the 
derivatives on the left-hand side do not exist on the space of all densities, 
for, if they existed, they would determine the potentials uniquely. 
Consequently, the functionals $T_s$ and $F$ display multiple derivative
discontinuities and must be redefined on equivalence classes of densities
arising from the potentials {\it modulo the nonunique pieces}.
The same applies to the xc functional $E_{xc}$ itself, 
since $E_{xc}$ is in general 
defined as the difference $E_{xc}= F -T_s -E_H$, where $E_H$ stands for all 
Hartree-like terms included in the respective formulation of DFT. 
Common approximations to $T_s$ and $E_{xc}$ do not display these
derivative discontinuities. Judging from experience with 
similar discontinuities in ordinary DFT \cite{levyperdew} we expect 
this shortcoming to be most relevant for the calculation of energy gaps.

The nonuniqueness problem discussed above (as well as the intimately 
related nondifferentiability problem) occurs, strictly speaking, only at zero 
temperature. At finite temperature one should work with a statistical ensemble,
rather than with a ground-state, and then the uniqueness of the relation 
between density and potential is restored \cite{argaman}.
However, the singularity at T=0 is an indicator that a real physical problem 
exists. Consider, as an illustration, the nonuniqueness of the potentials
of SDFT, discussed in Refs.~\cite{ep} and \cite{uniqueprl}.
Due to the nondifferentiability of the xc functional, an infinitesimal 
change in the spin density (such as the change $\delta m$ caused by the 
flipping of a single electron in an extended half-metallic ferromagnet)
may cause a finite (discontinuous) change in the xc potential. None of the 
existing approximations is able to reproduce such a discontinuity.
Going to finite but small temperatures simply replaces the discontinuity by a 
very rapid continuous change. 
To estimate the scale of this change we note that at $T=0$ the magnetic field
is only determined by the densities to within $E_G/\mu_0$, where $E_G$ is the 
energy gap and $\mu_0$ the Bohr magneton. Multiplying this with the 
low-temperature spin susceptibility we find that the spin density changes by
$(\delta m E_G/k_BT) exp (-E_G/k_BT)$.  As long as
$k_BT<<E_G$ this is much less than the physically relevant change
$\delta m$,  and therefore the functional remains effectively discontinuous
in the low-temperature regime.

In summary, we have shown that generalizations of DFT to current-carrying
and to superconducting systems suffer from the same nonuniqueness problem
we earlier discussed for the case of spin-polarized systems. 
Although the details are interestingly different in each of these three
cases, the physical connection of nonuniqueness with a vanishing
response function, as well as the classification of nonuniqueness
into systematic (arising from constants of motion) and accidental
(arising from special features of the ground state), and the consequences
for differentiability of the respective density functionals 
$T_s$ and $E_{xc}$ are the same in all three cases, and, we believe, also 
in any other generalization of DFT. 

{\bf Acknowledgments}
KC thanks the FAPESP for financial support.
GV acknowledges support from NSF Grant DMR-0074959.

\end{document}